\documentclass[aps,pre,floats,floatfix,superscriptaddress,nofootinbib,twocolumn]{revtex4}
\usepackage{latexsym,amsmath}
\usepackage{amsbsy}
\usepackage[usenames]{color}
\usepackage[pdftex]{graphicx}
\usepackage{amsfonts,amsmath,amssymb,hyperref}
\usepackage{color}
\usepackage{todonotes}
\usepackage{algorithm}
\usepackage{algorithmic}
\usepackage[usenames]{color}
\usepackage{bbm}
\usepackage{array}
\usepackage{multirow}

\begin{document}
\title{Stability of a Giant Connected Component in a Complex Network}

\author{Maksim Kitsak}
\affiliation{Department of Physics, Northeastern
University, 110 Forsyth Street, 111 Dana Research Center, Boston, MA, 02115, USA.}
\author{Alexander A. Ganin}
\affiliation{University of Virginia, Department of Systems and Information Engineering, Charlottesville, VA, 22904, USA}
\affiliation{U.S. Army Engineer Research and Development Center, Contractor, Concord, MA, 01742, USA}
\author{Daniel A. Eisenberg}
\affiliation{School of Sustainable Engineering and the Built Environment, Arizona State University, Tempe, AZ, 85281, USA}
\affiliation{U.S. Army Engineer Research and Development Center, Contractor, Concord, MA, 01742, USA}
\author{Pavel L. Krapivsky}
\affiliation{Department of Physics, Boston University, Boston, MA, 02215, USA}
\author{Dmitri Krioukov}
\affiliation{Departments of Physics, Mathematics, and Electrical and Computer Engineering, Northeastern University, Boston, MA, 02115, USA}
\author{David L. Alderson}
\affiliation{Operations Research Department, Naval Postgraduate School, Monterey, CA, 93943, USA}
\author{Igor Linkov}
\affiliation{U.S. Army Engineer Research and Development Center, Environmental Laboratory, Concord, MA, 01742, USA}
\date{\today}


\begin{abstract}
We analyze the stability of the network's giant connected component under impact of adverse events, which we model through the link percolation.  Specifically, we quantify the extent to which the largest connected component of a network consists of the same nodes, regardless of the specific set of deactivated links. Our results are intuitive in the case of single-layered systems: the presence of large degree nodes in a single-layered network ensures both its robustness and stability. In contrast, we find that interdependent networks that are robust to adverse events have unstable connected components. Our results bring novel insights to the design of resilient network topologies and the reinforcement of existing networked systems.
\end{abstract}

\maketitle

\begin{figure*}
\includegraphics[width=6.0in]{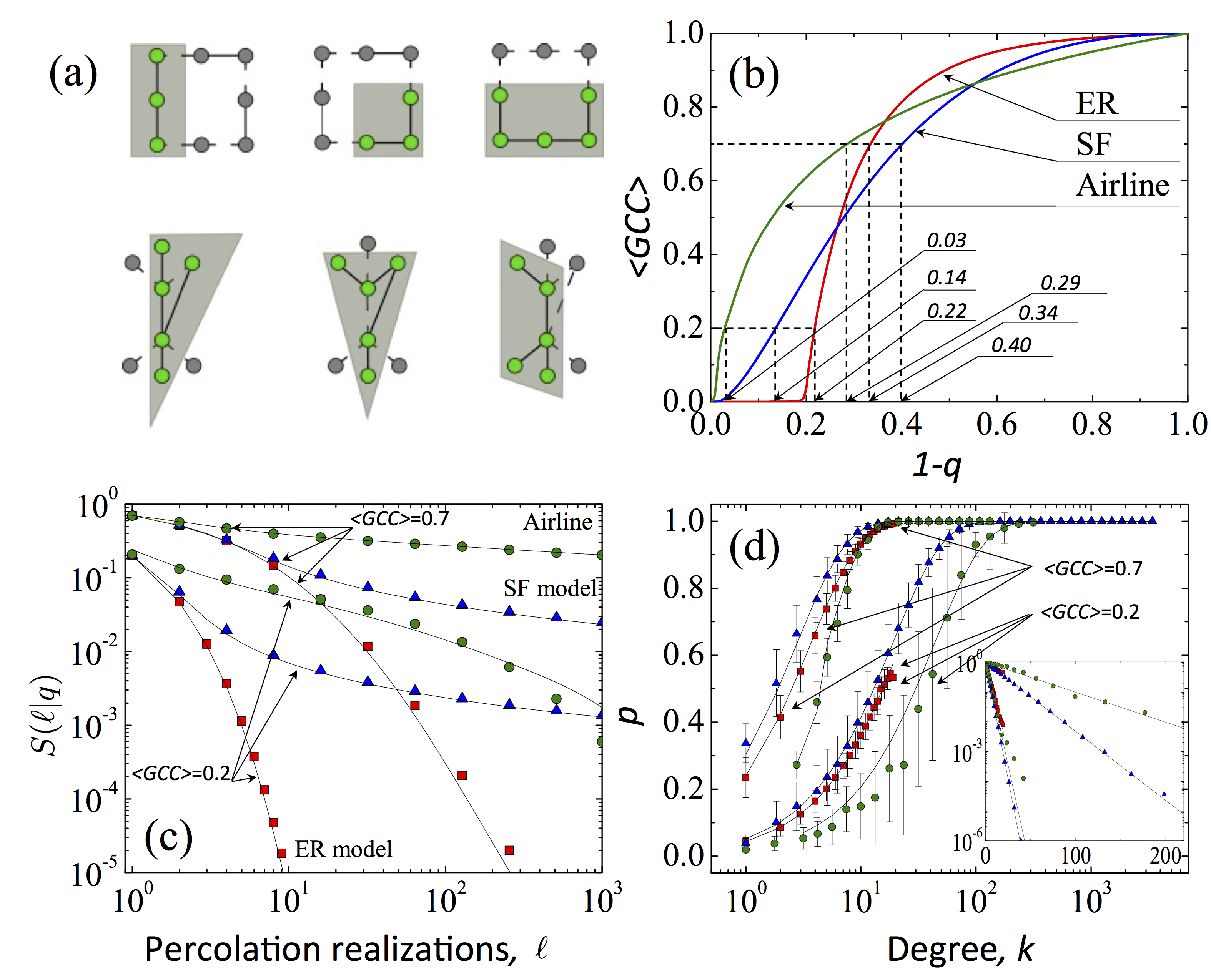}
\caption{\footnotesize   (a) GCCs shown for two toy networks at various links percolation realizations. Note that the
the GCC of the top network varies significantly depending on the set of removed links, while the GCC of the second network is more stable, and in all realizations includes the two central nodes. (b) The expected GCC size for  an ER and a SF models, as well as the airline network as a function of the fraction of deactivated links $q$. Both model networks are of the same size, $N=2\times10^{5}$ and the same average degree $\langle k \rangle=5.1$. The SF network is characterized by $\gamma = 2.5$ and minimum degree $k_{\rm min}=2$. Dashed lines show considered values of $q$ that are chosen such that $\langle GCC \rangle = 0.2$ and $\langle GCC \rangle = 0.7$. (c) Stability $S(\ell|q)$ as a function of $\ell$ for the (red squares) ER model, (blue triangles) SF model, and (green circles) the  Airline network described in Appendix~\ref{sec:realnets}. Solid lines correspond to analytical predictions by Eq.~(\ref{eq:persistence}). (d) Probability for a node to appear within the GCC of a network as a function of its degree. The color scheme is that of panel (c). Data is binned logarithmically, and error bars display the standard deviation values. The inset shows the probability of the node \emph{not} to appear within the GCC in the log-linear format with solid lines corresponding to analytical solutions given by Eqs.~(\ref{eq:qkinfty}) and (\ref{eq:u}).}
\label{fig:1}
\end{figure*}

\section{Introduction}

Robustness and resilience of networked systems under the impact of adverse events have been extensively studied in network science for two decades, but the research has been  primarily focused on computing the mean-field properties, such as the expected size of the system's giant connected component (GCC)~\cite{stauffer1994introduction,Dorogovtsev2010,cohen2010complex}.
The inherent assumption is that only connected subnetworks retain their functionality, with the largest of these being most relevant to the overall performance of the system.

While the mean-field analysis of a system's behavior is undoubtedly an important first step toward understanding its robustness, in most practical situations it is insufficient to know the expected size of the so-called `functional component.'
Rather, the location of the functional component within the network itself is important. It is especially true in the case of resilience where critical system function and its recovery is of prime importance ~\cite{ganin_operational_2016,linkov_resilience_2017}. For example, in the case of a major natural disaster, such as flood or an earthquake, one needs to know infrastructure units and transportation routes that are likely to remain functional. The efficiency  of immunization strategies depends on our knowledge of low and high-risk groups in social networks. Likewise,
the success of a marketing campaign depends on the knowledge of the target audience.

Yet, apart from a handful of recent works aiming to understand individual node properties in percolation and epidemic processes~ \cite{hamilton2014tight,morone2015influence,rogers2015assessing,kuehn2017heterogeneous,bianconi2017fluctuations}, organization patterns of individual network's functional components are poorly studied. One reason is that the random nature of adverse events coupled with the complexity of relevant networked systems often makes the prediction of functional subgraphs impossible. Indeed, consider two toy networks of the same size and average degree, which we repeatedly subject to adverse events of equal magnitude, modeled by link percolation, i.e., by deactivating a fraction of links selected uniformly at random, Fig.~\ref{fig:1}(a). Even though the expected GCC sizes of both networks are similar, the GCC of the first network is unstable and strongly depends on the set of deactivated links. In contrast, the GCC of the second network seems to be centered at the two largest degree nodes and only weakly depends on the set of deactivated links. Clearly, nodes of the first network are topologically identical and, thus, are equally likely to enter the GCC. The two central nodes of the second network, on the other hand, have much higher probability to enter the GCC compared to the remaining nodes, serving as anchors for its GCC.

Our work is motivated by two recent results in percolation theory~\cite{kuehn2017heterogeneous,bianconi2017fluctuations} analyzing heterogeneous network responses to different percolation realizations. In particular, Ref.~\cite{kuehn2017heterogeneous} demonstrates a considerable variation of network's connected component sizes and individual node probabilities to appear within the GCC. Concurrently, Ref.~\cite{bianconi2017fluctuations} introduces a framework to quantify GCC fluctuations by analyzing network responses to two random but possibly correlated percolation realizations. In our work we ask a related question: How stable is network's GCC? In other words, we aim to quantify the extent to which the giant connected component of a network consists of the same nodes, regardless of the specific set of deactivated links during an adverse event. We analyze both single and multi-layer networks, finding that large GCC size does not necessarily imply GCC stability.

\section{GCC stability of single-layer networks}

To simplify the exposition we model adverse events by link percolation~\cite{stauffer1994introduction,bollobas2006percolation} and limit our consideration to random network models with prescribed degree sequences. We also assume that there exists at most one GCC per network. At the same time we note that our approach is extendable to other types of percolation and more complex network topologies, including the situations with multiple GCCs per network~\cite{ben2008phase,chen2011explosive}. To quantify GCC stability we subject the network of interest to $\ell$ independent link percolation scenarios; each deactivating a random fraction $q$ of network links. We then compute the  sets of nodes $\Omega_{t}(q)$  for each percolation realization $t \in \{1,...,\ell\}$ and use them to define the GCC stability function as the fraction of nodes inside all sets $\Omega_{t}(q)$, namely
\begin{equation}
S(\ell|q) \equiv \frac{1}{N}\|\bigcap_{t=1}^{\ell}  \Omega_{t}(q) \|,
\end{equation}
where $N$ is the network size and $\| \Omega \|$ is the cardinality of set $\Omega$. In the case of GCCs consisting of random sets of nodes $S(\ell|q)$ decreases exponentially as a function of $\ell$, while in the case of stable GCCs $S(\ell|q)$ is expected to decrease slowly or remain constant. We first focus on GCC stability of the  Erd\H{o}s R\'{e}nyi (ER)~\cite{erdHos1960evolution} and scale-free (SF) random network models. The former is fully prescribed by the number of nodes $N$ and the constant connection probability $p$ for every node pair. The latter is characterized by the scale-free (SF) degree distributions, $p(k) \sim k^{-\gamma}$ and is generated by the Molloy-Reed algorithm~\cite{molloy1995critical}. Our results depicted in Fig.~\ref{fig:1}(b,c) indicate that GCC stability in ER is significantly smaller than that in SF networks. Indeed, $S(\ell|q)$ in the case of SF network models decays at significantly slower rates implying the existence of a stable sub-component, shared by all  $\Omega_{t}(q)$ sets. At the same time $S(\ell|q)$ in ER network models seems to decrease exponentially, albeit at rates slower than expected for GCCs consisting of randomly selected nodes.

Our results for $S(\ell|q)$ are consistent with the observation that individual node probabilities to belong to the GCC are strongly correlated with node's degree. As seen from Fig.~\ref{fig:1}(d), the probability for a node to belong to the GCC increases as a function of its degree and approaches $1$ for nodes of sufficiently large degree. These nodes belong to the GCC with high probability and serve as its anchors in a network, effectively stabilizing its GCCs. In contrast, the absence of such large degree anchor nodes in ER networks leads to unstable GCCs, Fig.~\ref{fig:1}(d).

To quantify the observed GCC stability phenomena we employ the generating function formalism~\cite{wilf2013generatingfunctionology}. The starting point of our analysis is the mean-field solution for the individual node probability to enter the GCC~\cite{Newman2010}:
\begin{equation}
p = 1 - \left[u(q)\right]^{k},
\label{eq:qkinfty}
\end{equation}
where $u(q)$  can be regarded as the mean-field probability that a given end of randomly chosen link leads to a connected component of finite size~\cite{callaway2000network}:
\begin{equation}
u = q + (1-q) G_{1}(u),
\label{eq:u}
\end{equation}
and $G_{1}(x)$ is the generating function for the outgoing edge degree distribution, $G_{1}(x) \equiv \sum_{k=0}^{\infty} p_{e}(k) x^{k}$, where $p_{e}(k) = (k+1) p(k+1)/\langle k \rangle$, and $p(k)$ is the degree distribution~\cite{newman2001random}.

Then, the probability that a node of degree $k$ is present in the GCC independently in $\ell$ percolation realizations is $\left(1-\left[u(q)\right]^{k}\right)^{\ell}$ and the expected stability of the GCC in $\ell$ independent percolation realizations is given by
\begin{equation}
S(\ell|q) = \sum_{k} p(k) \left(1-\left[u(q)\right]^{k}\right)^{\ell}.
\label{eq:persistence}
\end{equation}
To validate our theoretical results we solved Eq.~(\ref{eq:u}) numerically for each combination of $p(k)$ and percolation parameter $q$ used in Fig.~\ref{fig:1}(b). We then used the obtained $u$ values to superimpose the mean-field behaviors of individual node probabilities to enter the GCC and stability $S(\ell|q)$ with the numerical results, arriving at the excellent agreement between the two, Fig.~\ref{fig:1}(c,d).

The differences between $S(\ell|q)$ in SF and ER networks become more pronounced as expected GCC size decreases.
 Indeed, close to the criticality $S(\ell|q)$ in ER decays exponentially:
\begin{align}
S(\ell|q) &\approx \sum_{m=0}^{\ell} \binom{\ell}{m} (-1)^{m} \exp\left[-\langle k \rangle \left(1-  e^{-\epsilon \ell} \right) \right] \nonumber\\
& \sim \left[1-u^{\langle k \rangle}\right]^{\ell},
\end{align}
where $\epsilon \equiv 1-u \ll 1$ and $\langle k \rangle$ is the average degree in the ER network. In contrast, $S(\ell|q)$ in SF networks
\begin{equation}
S(\ell|q) \propto  \left(\epsilon k_0\right)^{\gamma-1} \int_{\epsilon k_{0}}^{\infty}\left[ 1- e^{-x}\right]^{\ell} \frac{{\rm d} x}{ x^{\gamma} } \sim \left(\frac{\epsilon k_0}{\ln \ell} \right)^{\gamma-1}
\end{equation}
for $\ell \gg 1$. Here $\gamma > 2$ is the degree distribution exponent, $p(k) \sim k^{-\gamma}$, see Appendix~\ref{sec:asymptotics} for derivation.
\begin{figure*}
\includegraphics[width=7in]{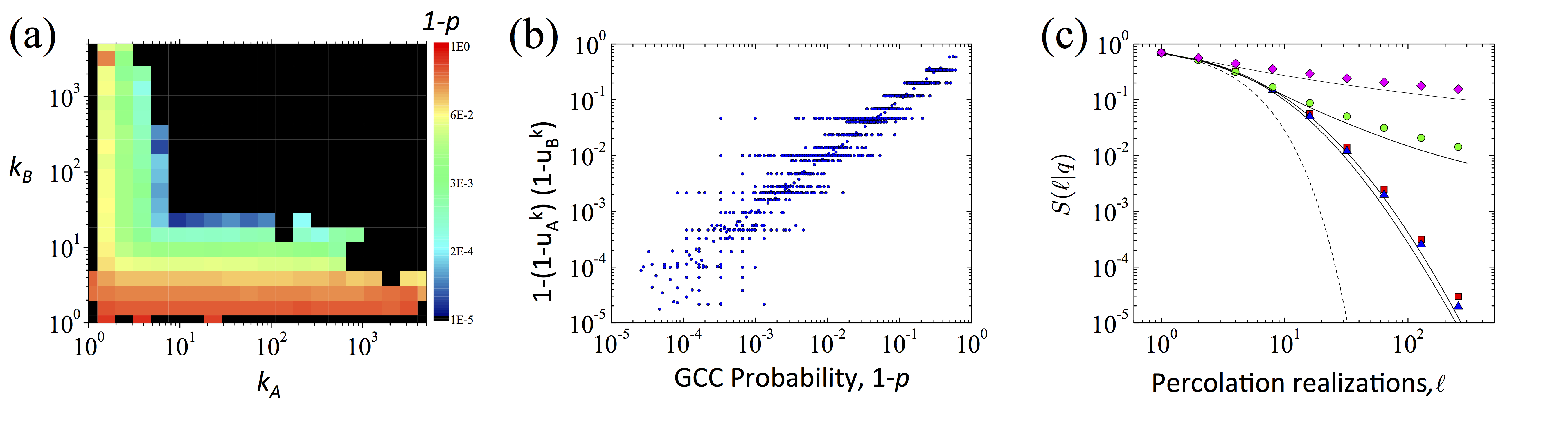}
\caption{\footnotesize  (a) Individual node probability to appear within the mutual GCC of the interconnected network as a function of its degrees in the network layers. The interdependent network consists of two SF layers with random interdependence links. Both SF layers are generated using parameters reported in Fig.~\ref{fig:1}. Link percolation thresholds are chosen as $q_{A}=0.49$ and $q_{B}=0$, corresponding to $\langle GCC \rangle=0.7$. (b) Individual node probabilities \emph{not} to appear within the GCC compared to the analytical estimate of Eq.~(\ref{eq:interdep_prob}). (c) Stability of the mutual GCC as a function of number of independent realizations $\ell$ calculated for (green circles) SF-SF, (blue triangles) ER-SF and (red squares) ER-ER interdependent network models  as well as (pink rhombi) the protein protein interaction network described in Appendix~\ref{sec:realnets}. Link percolation thresholds are, respectively,   $q_{A}=0.56$, $q_{A}=0.63$, $q_{A}=0.64$, and $q_A = 0.68$, $q_{B}=0$, all corresponding to $\langle GCC \rangle=0.7$. The dashed line corresponds to $S(\ell|q)$ of mutual GCCs consisting of random sets of nodes. Solid lines correspond to analytic solutions of Eqs~(\ref{eq:interdep_u}-\ref{eq:interdep_stab})}
\label{fig:3}
\end{figure*}

\section{GCC stability of multi-layer networks}

In the case of interdependent networks  the deactivation of nodes or links in one layer leads to the deactivation of nodes in other layers that depend on it~\cite{boccaletti2014structure}. In this case, the probability of a node to belong to the GCC depends not only on its own connections but also on the connections of supporter nodes in other layers. In the following, we consider a simple case of an interdependent network, consisting of equally sized layers $A$ and $B$ with one-to-one node interdependence; every layer representing a random network with prescribed degree distribution~\cite{buldyrev2009catastrophic}. Assuming no correlations in interdependencies between the two layers, the probability of a node to belong to the GCC is
\begin{equation}
p=\left(1-\left[u_{A}(\mathbf{q})\right]^{k_{A}}\right)\left(1-\left[u_{B}(\mathbf{q})\right]^{k_{B}}\right),
\label{eq:interdep_prob}
\end{equation}
where $k_{A}$ and $k_{B}$ are the degrees of the two interdependent nodes in layers $A$ and $B$, respectively, $\mathbf{q}\equiv \{q_{A},q_{B}\}$,  $q_{A}$ and $q_{B}$ are the initial fractions of deactivated links in layers $A$ and $B$ respectively; while parameter $u_{A}$ ($u_{B}$) is the effective probability that a given end of a randomly chosen link in layer $A$  ($B$) leads to a mutual connected component of finite size. The mean-field parameters $u_A$ and $u_B$ can be found from the recursive approach of Ref.~\cite{buldyrev2009catastrophic} and are given by the system of equations
\begin{eqnarray}
\label{eq:interdep_u}
1-u_A &=& \left(1-q_{A}\right)\left(1 - G_{1}^{A}(u_A)\right)\left(1 - G_{0}^{B}(u_B)\right),  \\
\label{eq:interdep_u2}
1-u_B &=& \left(1-q_{B}\right)\left(1 - G_{1}^{B}(u_B)\right)\left(1 - G_{0}^{A}(u_A)\right),
\end{eqnarray}
where $G_{\{0,1\}}^{\{A,B\}}(x)$ are the generation functions for the regular and the outgoing degree distributions of layers $A$ and $B$, see Fig.~\ref{fig:3}(a,b) and Appendix~\ref{sec:interstab} for derivation. We note that this result is in good agreement with numerical results, which we obtain for ER-ER, ER-SF and SF-SF models as well as the network of protein-protein interactions (PPI), and depict for SF-SF network in  Fig.~\ref{fig:3}(a).

Since interdependence links between the layers are established randomly, the joint degree distribution of the network is $p(k_A,k_B) = p_{A}(k_A)p_{B}(k_B)$, where  $p_{A,B}(k)$ are degree distributions of layers $A$ and $B$ respectively. Then, the stability of the mutual connected component is simply the product of the contributions from the two layers:
\begin{equation}
S_{AB}(\ell|\mathbf{q})= S_{A}(\ell|\mathbf{q})S_{B}(\ell|\mathbf{q}),
\label{eq:interdep_stab}
\end{equation}
where $S_{A}$ and $S_{B}$ are the stability contributions of domains $A$ and $B$ respectively: $S_{A} \equiv \sum_{k} p_{A}(k) \left(1 - \left[u_{A}(\mathbf{q})\right]^{k}\right)^{\ell}$, and $S_{B}$ is defined similarly. This result together with our results in Fig.~\ref{fig:1} for single layered networks explains
stable mutual GCCs in SF-SF networks and unstable GCCs in ER-ER and ER-SF networks that we observe in Fig.~\ref{fig:3}(c).

\section{Conclusion}

 Taken together, our results indicate that networks with broader degree distributions are characterized by stable GCCs. Large degree nodes in these networks are part of the GCC with probabilities close to $1$, and can be regarded as anchors keeping the GCC in place.
 At the same time, it is important to note that network robustness and GCC stability do not in general imply each other. We find that
 interdependent networks with broad degree distribution of their layers (e.g., SF-SF networks) have stable mutual GCCs. At the same time, however,
 SF-SF networks are known to be vulnerable to cascading failures~\cite{buldyrev2009catastrophic}. Indeed, large degree nodes in interdependent networks are likely to depend on small degree nodes in other layers. As a result, failures of these small degree nodes immediately lead to failures of the large degree nodes that depend on them. In contrast, interdependent networks with narrow degree distributions, e.g., ER-ER networks, are less vulnerable due to a more uniform mixing of node degrees in their layers. The stability of connected components in interdependent networks, on the other hand, exclusively relies on the existence of anchor nodes, which according to Eq.~(\ref{eq:interdep_prob}) are large degree nodes dependent on other large degree nodes. While these anchor nodes are not frequent in SF-SF networks, they are non-existent in ER-ER networks, explaining why mutual GCCs are not stable in the latter.

 Our findings open new  avenues toward the design of efficient network reinforcement strategies. Indeed, building upon our results one can formulate the subgraph reinforcement problem as an optimization. Similar to the GCC stability, one can define the stability of any subgraph $\Omega$ of interest:
\begin{equation}
S_{\Omega}(\ell|q) \equiv \|\bigcap_{t=1}^{\ell}  \Omega_{t}(q)\|=  \sum_{i \in \Omega} \left(1- \left[u(q)\right]^{k_{i}}\right)^{\ell},
\end{equation}
where $u$ is the mean-field probability that a given end of a randomly chosen link leads to a cluster of finite size and is given by Eq.~(\ref{eq:u}).
Then the reinforcement strategy is defined in using extra resources, e.g. additional nodes or links to maximize the $S_{\Omega}(\ell|q)$ objective function. It is also immediately clear that the optimal reinforcement strategies depend on the number of expected adverse events $\ell$:  if the system is designed to withstand a single adverse event, the reinforcement strategy should be aimed at maximizing the expected GCC size. On the other hand, if the system is designed to withstand multiple adverse events, the investments should be made to further reinforce or create the anchor nodes.

The mean-field analysis offered here has important limitations. The generating function approach works well for locally tree-like networks, while real systems contain significant amount of short loops. Likewise, it is now understood that heterogeneous distributions in particular can have considerable diversity in their topologies~\cite{AldersonLi2007} and that the hubs in these networks are not always centrally located. As a result, we do notice significant deviations for individual node probabilities in real networks, Fig.~\ref{fig:1}(d), that can't be explained by the mean-field approach.

Finally, within the engineering sciences there is a general understanding that the organization of many real-world networks is governed by processes that are not sufficiently characterized by stylized models of random networks and that these differences can have important implications for understanding robustness and vulnerability~\cite{doyle2005robust,florin_irgc_2016}.
There is a growing body of work in large-scale optimization and game theory that leverages the specialized structure of these networks to discover specific vulnerabilities and to identify limited defensive investments that can maximally increase robustness and resilience (e.g., Ref.~\cite{alderson2015operational,ganin_operational_2016}). Bridging the gap between the specific recommendations of these highly detailed models and the insights from more general models described here is an important goal for ongoing research.

\section{Acknowledgements}

We thank Shlomo Havlin and Pim van der Hoorn for useful discussions and suggestions. This work was supported by the U.S.~Defense Threat Reduction Agency grant DTRA-18681-M. MK and DK were additionally supported by NSF grants CCF-1212778,  CCF-1212778, and IIS-1741355, and ARO grants W911NF-16-1-0391 and W911NF-17-1-0491. The views and opinions expressed in this article are those of the individual authors and not those of the U.S. Army, or other sponsor organizations.

%

\appendix

\section{Real networks}
\label{sec:realnets}
We test our GCC stability results on two real networks. The first real network is the complete US airport network in 2010, where nodes are US airports and an undirected link is established between two airports if there is a commercial flight between them. The network consists of $N=1,574$ nodes and  $E=28,236$ links, and is publicly available from \url{https://openflights.org/} and \url{https://toreopsahl.com/datasets/}.

The second is the network of human protein-protein interactions (PPI) that we derived from the BioGRID database~\cite{Biogrid}. We represent the PPI dataset as a two-layer interdependent network, where layer $A$ is formed by direct protein interactions, and links in layer $B$ correspond to physical associations between proteins. In our analysis we focus on the mutual connected component of the PPI network that contains $N=11,365$ nodes in each layer with the average degrees of $\langle k_A \rangle = 11.00$ and $\langle k_B \rangle = 25.38$ respectively.

\section{GCC stability of single layer networks close to the criticality}
\label{sec:asymptotics}
Here we consider the asymptotic behavior of $S(\ell|q)$ for three types of network models: regular, ER, and SF network models.

 \subsection{Regular network models}

 All nodes of a regular network model have the same degree $k_{0}$, resulting in $p(k) = \delta(k,k_{0})$, where $\delta(m,n)$ is the Kronecker delta. Then, GCC stability of a regular network is trivially $S(\ell|q) = \left( 1 - u(q)^{k_{0}}\right)^{\ell}$ for all $u(q)$ values.
 \subsection{ER network models}
 ER network models are characterized by the Poisson degree distribution, $p(k) = e^{-\langle k \rangle} \langle k \rangle^{k}/k!$, where $\langle k \rangle = p(N-1)$ is the average degree. In order to obtain the asymptotic expression for $S(\ell|q)$  we first expand $\left( 1 - u(q)^{k}\right)^\ell$ into the binomial sum. Then by reversing the summation order in Eq.~(\ref{eq:persistence}) we obtain
\begin{equation}
S(\ell|q) = \sum_{m=0}^{\ell} \binom{\ell}{m} (-1)^{m} \exp\left( - \langle k \rangle \left( 1 - u(q)^{m}\right) \right).
\label{eq:pers_er}
\end{equation}
Close to the criticality  $\epsilon(q)\equiv 1-u(q)\ll 1$ and therefore, $1 - u(q)^{m} \simeq \epsilon(q) m$. Plugging this into
Eq.~(\ref{eq:pers_er}) results in
\begin{equation}
S(\ell|q) \approx \left[1-e^{-\epsilon(q) \langle k \rangle}\right]^{\ell} \approx \left[1-u(q)^{\langle k \rangle}\right]^{\ell}.
\end{equation}

\subsection{SF network models}

SF network models are characterized by power-law degree distributions $p(k) \approx (\gamma-1) k_{0}^{\gamma -1} k^{-\gamma}$,
where $k_{0}$ is the smallest degree and $\gamma > 2$.

To deduce the asymptotic behavior of $S(\ell|q)$ we first approximate $u(q)^{k}$ in Eq.~(\ref{eq:persistence}) as $e^{- k\epsilon(q)}$, where $\epsilon(q) \equiv 1- u(q) \ll 1$. Then, by approximating the summation in Eq.~(\ref{eq:persistence}) with an integral we obtain:
\begin{eqnarray}
S(\ell|q) &\approx& (\gamma -1) \left[\epsilon(q) k_{0}\right]^{\gamma -1} I_{\ell}(\gamma),\\
 I_{\ell}(\gamma) &\equiv& \int_{\epsilon(q)k_{0}}^{\infty} x^{-\gamma} \left[1 - e^{-x} \right]^{\ell} {\rm d} x.
\label{Int_asymp}
\end{eqnarray}
Further, to deduce the asymptotic behavior of (\ref{Int_asymp}) we note that the function $\left[1-e^{-x}\right]^\ell$ is monotonically increasing from 0 to 1 on the interval $0<x<\infty$. It is useful to think about $\left[1-e^{-x}\right]^\ell$ as a smoothed version of the step function $\theta(x-\ln \ell)$. To appreciate the shift we center the function $\left[1-e^{-x}\right]^\ell$ at the position $x_0$ where it equals $e^{-1}$. From $\left[1-e^{-x_0}\right]^\ell=e^{-1}$ we obtain $e^{-x_0}=1-e^{-1/\ell}$ leading, when $\ell\gg 1$, to
\begin{equation}
x_0 = \ln \ell
\end{equation}
and thereby explaining the $\ln \ell$ shift. To appreciate the width of the quasi-step function $\left[1-e^{-x}\right]^\ell$ we define the left boundary through $[1-e^{-x_-}]^\ell=e^{-M}$ and the right boundary through $[1-e^{-x_+}]^\ell=1-e^{-M}$. These definitions are reasonable if we choose $M\gg 1$. From these definitions
\begin{equation}
\label{x:pm}
x_- = \ln \ell - \ln M, \qquad  x_+ = \ln \ell + M
\end{equation}
Thus the quasi-step function $[1-e^{-x}]^\ell$ has the following properties:
\begin{enumerate}
\item It is essentially vanishes when $0<x<x_-$.
\item It monotonically increases from 0 to 1 in the interval $x_-<x<x_+$.
\item It is essentially equal to 1 in the interval $x>x_+$.
\end{enumerate}

\begin{figure*}
\includegraphics[width=7in]{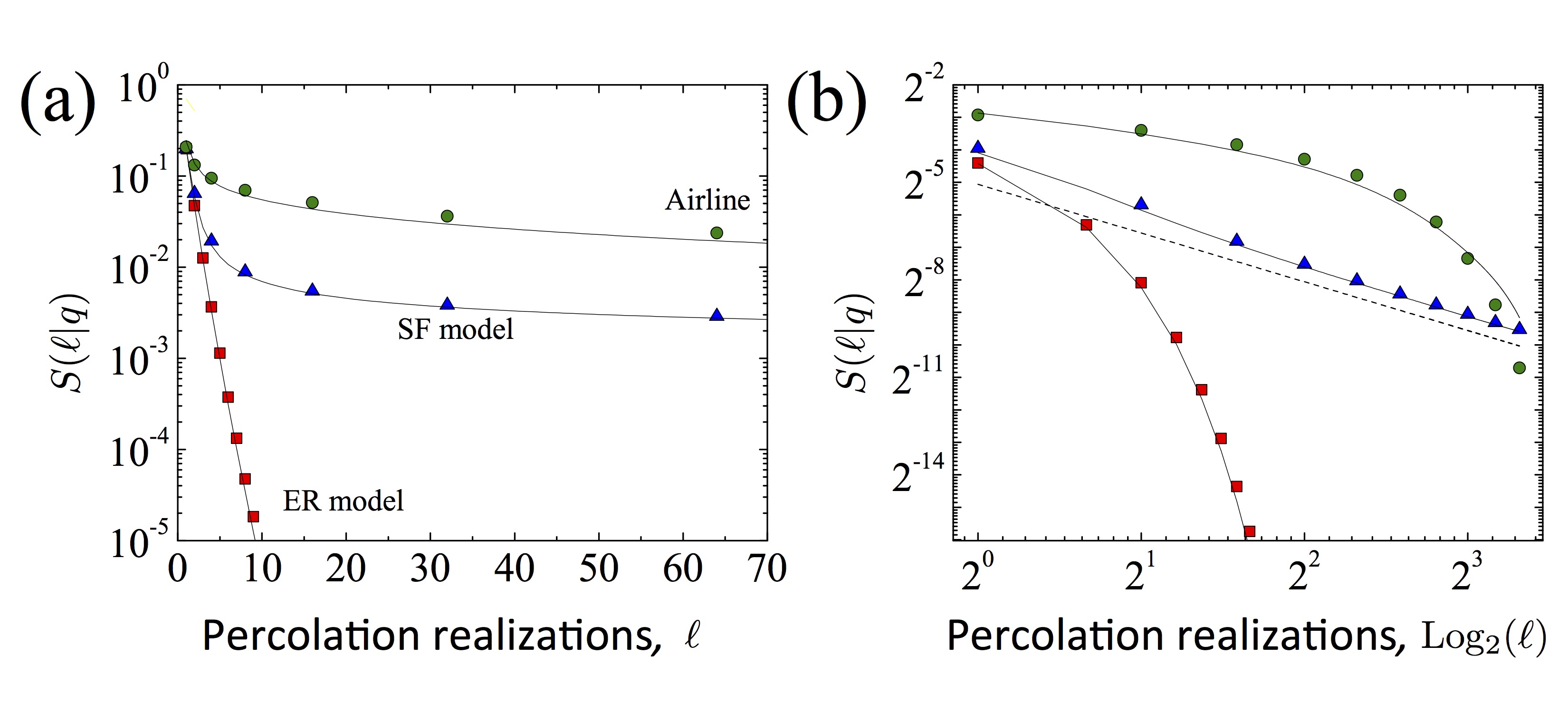}
\caption{\footnotesize  Stability $S(\ell|q)$ as a function of (a) $\ell$  and (b) ${\rm} Log_{2} (\ell)$ shown at $\langle GCC \rangle = 0.2$ for the (red squares) ER model, (blue triangles) SF model, and (green circles) the Airline network. All parameters are those of Fig.~1(b,c) in the main text. The dashed line in panel (b) corresponds to the asymptotic regime of $S(\ell|q) \sim \left[{\rm Ln}~\ell \right]^{-1.5}$ prescribed by Eq.~(6).}
\label{fig:SI}
\end{figure*}

We next establish the upper and lower bounds for the integral in (\ref{Int_asymp}). The upper bound is obtained by replacing $\left[1-e^{-x}\right]^\ell$ by unity when $x>x_-$ and $e^{-M}$ when $k_0\epsilon(q)<x<x_-$; the lower bound is obtained by replacing $\left[1-e^{-x}\right]^\ell$ by $1-e^{-M}$ when $x>x_+$ and zero when $0<x<x_+$. Since $\int_{x_*}^\infty \frac{dx}{x^\gamma}= \frac{1}{\gamma-1}\,\frac{1}{x_*^{\gamma-1}}$, the bounds are
\begin{equation}
\label{Int:bounds}
\frac{1-e^{-M}}{x_+^{\gamma-1}} \leq (\gamma - 1)I_\ell(\gamma) \leq \left(\frac{1}{x_-^{\gamma-1}} + e^{-M}\frac{1}{\left[\epsilon(q)k_0\right]^{\gamma-1}} \right)
\end{equation}
Since both $x_\pm \to \ln \ell$ in the $\{\ell, M\}\to\infty$ limit, such that $1 \ll M \ll \ell$, $I_\ell(\gamma) \sim \ln \ell ^{1-\gamma}$ and
\begin{equation}
P(\ell|q) \sim \left[\frac{k_{0} \epsilon(q)}{ \ln \ell}\right]^{1-\gamma},
\end{equation}
see Fig.~\ref{fig:SI}.

\section{GCC stability of interdependent networks}
\label{sec:interstab}
Link percolation results for interdependent networks can be recovered from Ref.~\cite{buldyrev2009catastrophic} by setting the fraction of active nodes to 1 and replacing degree generating functions with those for link percolated networks. Indeed, let us introduce the generating functions for degree distributions of layers $A$ and $B$ as
\begin{eqnarray}
G_{0}^{\{A,B\}}(x) &\equiv& \sum_{k} p_{\{A,B\}}(k) x^{k}, \\
G_{1}^{\{A,B\}}(x) &\equiv& \sum_{k}\frac{(k+1)}{{\langle k_{\{A,B\}} \rangle}} p_{\{A,B\}}(k+1)x^{k},
\end{eqnarray}
where $p_{A}(k)$ and $p_{B}(k)$ are the degree distributions of domains $A$ and $B$ respectively, while  $\langle k_{A} \rangle$ and $\langle k_{B} \rangle$ are the corresponding average degree values.

Then the expected size of the mutual connected component $\mu_{\infty}$ is given by
\begin{equation}
\mu_{\infty} = xy,
\end{equation}
where the mean-field parameters $x$ and $y$ can be regarded as the effective fractions of nodes belonging to the GCC of layers $A$ and $B$ respectively, and are given by the system of four equations:
\begin{eqnarray}
\label{eq:x}
x& =& 1-G_{0}^{A}\left( 1-y(1-f_{A})\right),\\
\label{eq:y}
y&=&1-G_{0}^{B}\left( 1-x(1-f_{B})\right),\\
\label{eq:fa}
f_{A} &=& G_{1}^{A}\left( 1-y(1-f_{A})\right),\\
\label{eq:fb}
f_{B} &=& G_{1}^{B}\left( 1-x(1-f_{B})\right).
\end{eqnarray}
Here parameters $f_{A}$ and $f_{B}$ in analogy with the single layer case can be regarded as the effective probabilities that a given end of a
randomly chosen link in layer $A$ or $B$, leads to a connected component of a finite size.

By making use of Eqs.~(\ref{eq:x}) and (\ref{eq:y}) one can rewrite the expression for the mutual connected component as
\begin{equation}
\mu_{\infty} = \left(1-G_{0}^{A}\left( 1-y(1-f_{A})\right) \right) \left(1-G_{0}^{B}\left( 1-x(1-f_{B})\right)\right),
\end{equation}
from which one can immediately extract the individual node probability to appear within the mutual GCC:
\begin{equation}
p=\left(1 - u_{A}^{k_{A}}\right) \left( 1 - u_{B}^{k_{B}} \right),
\end{equation}
where $k_{A}$ and $k_{B}$ are the degrees of the two interdependent nodes in layers $A$ and $B$, and the mean-field parameters $u_{A}$ and $u_{B}$ are given by
\begin{eqnarray}
u_{A}=1-y\left(1-f_{A}\right),\\
u_{B}=1-x\left(1-f_{B}\right).
\end{eqnarray}

The same equations hold in the link percolation case, except the original generating functions $G_{\{0,1\}}^{\{A,B\}}(x)$ need to be replaced with those for percolated layers:
\begin{eqnarray}
\tilde{G}_{\{0,1\}}^{A}(x) &=& G_{\{0,1\}}^{A}\left(1-\left(1-q_A\right)(1-x)\right), \\
\tilde{G}_{\{0,1\}}^{B}(x) &=& G_{\{0,1\}}^{B}\left(1-\left(1-q_B\right)(1-x)\right),
\end{eqnarray}
where $q_{A}$ and $q_{B}$ are the fractions of initially deactivated links in layers $A$ and $B$ respectively.

After a series of straightforward substitutions and simplifications the final result reads
\begin{equation}
p = \left(1-\left[u_{A}(\mathbf{q})\right]^{k_{A}}\right)\left(1-\left[u_{B}(\mathbf{q})\right]^{k_{B}}\right),
\end{equation}
where
\begin{eqnarray}
1-u_A &=& \left(1-q_{A}\right)\left(1 - G_{1}^{A}(u_A)\right)\left(1 - G_{0}^{B}(u_B)\right), \nonumber \\
1-u_B &=& \left(1-q_{B}\right)\left(1 - G_{1}^{B}(u_B)\right)\left(1 - G_{0}^{A}(u_A)\right), \nonumber
\end{eqnarray}
reproducing Eqs.~(\ref{eq:interdep_u}) and (\ref{eq:interdep_u2}) in the main text.

\end{document}